\documentstyle{mn}

\newcommand{\eg}{{\sl e.g. }}
\newcommand{\ie}{{\sl i.e. }}
\newcommand{\etal}{{\sl et al. }}

\newcommand{\Msun}{\mbox{$M_{\odot}$}}

\newcommand{\ltsimeq}{\raisebox{-0.6ex}{$\,\stackrel 
        {\raisebox{-.2ex}{$\textstyle <$}}{\sim}\,$}} 
\newcommand{\gtsimeq}{\raisebox{-0.6ex}{$\,\stackrel
        {\raisebox{-.2ex}{$\textstyle >$}}{\sim}\,$}}

\newcommand{\JHK}{$J\!H\!K~$}
\newcommand{\degs}{$^{\circ}$}

\newcommand{\rod}{Rodr\'\i guez }

\title[Multi-Wavelength Monitoring of GRS 1915+105]
{Multi-Wavelength Monitoring of GRS 1915+105} 
\author[Bandyopadhyay et al.]{
R. Bandyopadhyay,$^{1}$ P. Martini,$^{2}$ E. Gerard,$^{3}$ P.A. Charles,$^{1}$ R.M. Wagner,$^{2}$ \and C. Shrader,$^{4}$ T. Shahbaz,$^{1}$ and I.F. Mirabel$^{5,6}$ \\ 
$^{1}$University of Oxford, Department of Astrophysics,
Nuclear Physics Building, Keble Road, Oxford, OX1 3RH, UK \\
$^{2}$Department of Astronomy, Ohio State University, 174 West 18th Avenue, Columbus, OH 43210-1106, USA \\
$^{3}$Dept. ARPEGES, Observatoire de Paris, Section de Meudon, 5 Place Jules Janssen, 92195 Meudon, France \\
$^{4}$NASA/Goddard Space Flight Center, Greenbelt, MD 20771, USA \\
$^{5}$Service d'Astrophysique/CEA, DSM/DAPNIA, Centre d'Etudes de Saclay, F-91191 Gifsur-Yvette, France \\
$^{6}$Instituto de Astronomia y Fisica del Espacio, Argentina 
}

\begin{document}

\maketitle

\begin{abstract}
Since its discovery in 1992, the superluminal X-ray transient GRS 1915+105 has 
been extensively observed in an attempt to understand its behaviour.  We 
present here first results from a multi-wavelength campaign undertaken 
from July to September 1996.  This study includes X-ray data from the RXTE All 
Sky Monitor and BATSE, two-frequency data from the Nancay radio telescope, and 
infrared photometry from the 1.8m Perkins telescope at Lowell Observatory.  
The first long-term well-sampled IR light curve of GRS 1915+105 is presented 
herein and is consistent with the interpretation of this source as a 
long-period binary.  We compare the various light curves, searching 
for correlations in the behaviour of the source at differing wavelengths and 
for possible periodicities.  

\noindent

\end{abstract}

\section{Introduction}
The X-ray transient GRS 1915+105 was discovered by the GRANAT satellite in 1992 (Castro-Tirado \etal 1994).  It has since shown erratic activity with recurrent rises to a maximum X-ray luminosity of 3x$10^{39}$ ergs/s (Chaty \etal 1996, hereafter C96).  VLA observations at the time of outburst led to the discovery of relativistic ejections of plasma clouds with apparent superluminal motions, possibly a smaller-scale analogue to the jets observed in active galactic nuclei and quasars (Mirabel \& \rod 1994, hereafter MR94).  GRS 1915+105 has a highly variable spectral index, a hard X-ray tail at $\geq$150 keV, and often exceeds the Eddington luminosity limit for a neutron star.  Due to these characteristics and its similarity to the other Galactic superluminal source, GRO J1655-40 (with a compact object mass of 7\Msun; Orosz \& Bailyn 1997), GRS 1915+105 is a suspected black hole binary.  Radio observations show that it is at a kinematic distance $D$ = 12.5$\pm$1.5 kpc from the Sun; combined with measurements of the hydrogen column density along the line of sight, a visual extinction of $A_{V}$ = 26.5$\pm$1 mag has been determined (C96).  Variable radio and infrared counterparts to the X-ray source have been found (Mirabel \etal 1994).  However, the optical counterpart has only been detected at $I$, with a magnitude of 23.4; upper limits on the $B$, $V$, and $R$ magnitude are 25.9, 26.1, and 26.1 respectively (Boer, Greiner, \& Motch 1996).   

The extremely high visual extinction ensures that GRS 1915+105 is impossible to study optically.  Therefore, in an attempt to constrain the nature of the binary system, an ongoing program of infrared observations of GRS 1915+105 is currently underway.  Of particular interest is the nature of the mass-donating star.  On the basis of its spectral morphology and the calculated absolute $K$ magnitude, Castro-Tirado \etal (1996, hereafter CT96) suggested GRS 1915+105 to be a low-mass X-ray binary (LMXB).  However, further $K$-band spectroscopy of the counterpart revealed several prominent emission features but no detectable absorption features which would be indicative of a late-type secondary.  On the contrary, the $K$-band spectrum is very similar to those of high-mass X-ray binaries (HMXB) with a Be star as the mass-losing component (Mirabel \etal 1997, hereafter M97).  In addition, M97 show that CT96 underestimated the visual extinction along the line of sight to GRS 1915+105.  Using our corrected $A_{V}$, a striking similarity becomes apparent between the observed absolute IR magnitudes and colours of GRS 1915+105 and those of other HMXBs, including SS433 (the classic stellar source of relativistic jets; Margon 1984) and A0538-66 (Allen 1984).

\JHK photometry of the counterpart has revealed both short- and long-term variability; changes of one magnitude over an interval of 24 hours and two magnitudes over several months have been observed, but no periodicity has been seen (C96).  However, all previous IR photometry of GRS 1915+105 has been either short term, spanning only a few days, or highly erratic, with a handful of observations scattered across an extended time period.  In order to search for the long-term periodicity ($\sim$15-45 days) that would be expected in an eccentric HMXB, a high-resolution ($\sim$1 day) light curve with a baseline of several months is necessary.  To this end, we observed GRS 1915+105 in the $K$-band from July to September 1996, producing the first long-term IR light curve of this source.

Since the discovery of the superluminal ejections, GRS 1915+105 has been under almost continuous surveillance at X-ray and radio wavelengths.  The Rossi X-ray Timing Explorer (RXTE) and the Burst and Transient Source Experiment (BATSE) on the Compton Gamma Ray Observatory (CGRO) have monitored the X-ray behaviour of GRS 1915+105 extensively (see \eg Greiner \etal 1996, hereafter G96, and Harmon \etal 1996), while numerous radio observations have been performed using the Very Large Array (VLA), Nancay Telescope, Green Bank Interferometer (GBI), and Ryle Telescope (\rod \& Mirabel 1997; \rod \etal 1995; Pooley 1996).  At these wavelengths, GRS 1915+105 shows fluctuations of variable amplitude on a variety of timescales.  As in the IR, no consistent periodicity has been found.  However, some X-ray and radio observations have shown a quasi-periodicity of $\sim$30 days (\rod \& Mirabel 1995).  Numerous attempts have been made to correlate the X-ray and radio activity of GRS 1915+105, but the wide range of behaviour exhibited by the source has made such comparison complex.  

Here we present multi-wavelength light curves of GRS 1915+105 from the period July to September 1996, including $K$-band photometry from the 1.8m telescope at Lowell Observatory, RXTE 2-10 keV, BATSE 20-200keV, and 1414 MHz and 3310 MHz observations from the Nancay radio telescope.  By exploring the source's behaviour simultaneously at multiple wavelengths, we attempt to place some constraints on the nature of the system.  In addition, we discuss possible interpretations for the correlations (or lack thereof) between the light curves.  This study is the first to include extensive IR photometry as well as X-ray and radio information in a multi-wavelength study of GRS 1915+105.

\section{Observations and Data Reduction} 

\subsection{Infrared}

Our $K$-band images were obtained over a total of 28 nights during observing runs in 1996 July, August, and September using the Ohio State Infrared Imager/Spectrometer (OSIRIS; DePoy et al. 1993) on the Perkins 1.8m telescope of the Ohio State and Ohio Wesleyan Universities at Lowell Observatory.  The images were taken with the f/7 camera which provides a 2\farcm7 field of view at a resolution of 0\farcs63 pixel$^{-1}$; our seeing was typically 1\farcs8.  On each night, a series of five 30-second (coadded) exposures were obtained, with offsets of $\sim20''$ between each exposure.  The images were processed by first running a simple interpolation program to remove bad pixels.  A median-combined sky image was then created from the five exposures and subtracted from each image.  The resulting sky-subtracted frames were then flat-fielded with an image constructed from the difference of dome flats obtained with the flat-field lamps on and off.  

The data were analyzed using the IRAF reduction task ``apphot'' with a 4.5 pixel aperture.  The aperture was chosen to maximize the target signal relative to the sky background; it also excludes a star 4'' northeast of the target.  Relative photometry was then performed using several field standards.  The July-August monsoon season caused recurrent poor weather during the three month run, producing large photometric errors on affected nights, on the order of $\pm$0.3 magnitudes; however, more typical errors are $\pm$0.08 mag.  On the night of 6 August, absolute photometry of the local standards was obtained and used to calculate the $K$ magnitudes of GRS 1915+105 throughout the three-month light curve.  A journal of the IR observations with $K$ magnitudes is presented in Table 1.

\begin{table}
\caption{$K$ magnitudes of GRS 1915+105}
\begin{center}
\begin{tabular}{lccl}\hline
UT Date & K magnitude & \# of images & Comments \\ 
1996 		& 	& (sets of 5) & \\ \hline			
2/7 &  12.42$\pm$0.05 & 2 &  \\ 
3/7 &  12.37$\pm$0.09 & 3 &  \\ 
5/7 &  13.09$\pm$0.30 & 1 & mostly cloudy \\ 
8/7 &  13.37$\pm$0.25 & 2 & mostly cloudy \\ 
11/7 &  13.11$\pm$0.04 & 1 & \\ 
12/7 &  13.49$\pm$0.12 & 1 & variable clouds\\ 
13/7 &  13.29$\pm$0.15 & 1 & variable clouds\\ 
14/7 &  13.43$\pm$0.08 & 2 &  \\ 
15/7 &  13.48$\pm$0.04 & 2 &  \\ 
6/8 &  12.95$\pm$0.06 & 1 &  \\ 
7/8 &  12.76$\pm$0.20 & 1 & poor weather \\ 
8/8 &  13.14$\pm$0.06 & 1 & \\ 
10/8 &  13.00$\pm$0.09 & 1 &  \\ 
11/8 &  13.20$\pm$0.08 & 1 &  \\ 
12/8 &  13.14$\pm$0.05 & 1 &  \\ 
13/8 &  13.23$\pm$0.09 & 1 &  \\ 
14/8 &  13.19$\pm$0.10 & 1 &  \\ 
16/8 &  13.45$\pm$0.18 & 1 & poor weather \\ 
20/8 &  13.48$\pm$0.14 & 3 &  \\ 
27/8 &  13.37$\pm$0.09 & 1 &  \\ 
4/9 &  13.51$\pm$0.08 & 1 &  \\ 
8/9 &  13.72$\pm$0.05 & 1 &  \\ 
9/9 &  13.47$\pm$0.15 & 1 &  \\ 
15/9 &  12.85$\pm$0.06 & 2 &  \\ 
16/9 &  13.17$\pm$0.02 & 1 &  \\ 
20/9 &  12.93$\pm$0.38 & 1 & poor weather \\ 
22/9 &  12.99$\pm$0.18 & 1 & poor weather \\ 
23/9 &  13.00$\pm$0.02 & 1 &  \\ \hline
\end{tabular}
\end{center}
\end{table}	

\subsection{Radio}

Radio monitoring of GRS 1915+105 was initiated with the Nancay radio telescope in late 1993 (\rod \etal 1995, hereafter R95) at 21 cm and 9 cm wavelengths.  It was interrupted in 1995 and resumed in May 1996 with some improvements.  In order to minimize any possible outside interference, the observations were carried out at 1404, 1410, 1416 and 1424 MHz (21 cm) and 3263, 3290, 3335 and 3349 MHz (9 cm), all with 6.4 MHz bandwidths falling in protected radio astronomy bands of HI and CH, save for 3290 MHz.  Each daily run consisted of averaging RA drift scans centered on the source.  The Nancay radio telescope has half-power beamwidths of 4' x 22' at 21 cm and 2' x 10' at 9 cm.  The confusion level due to nearby sources is usually 20 mJy and 5 mJy, respectively, but due to the proximity of the Galactic plane, it rises to 400 mJy and 100 mJy in the vicinity of GRS 1915+105.  It is thus essential to observe the source at the same hour angles so that the antenna beam profile remains constant. 

The flux density values are obtained by subtracting a ``reference'' drift scan or template from each daily average drift scan and fitting a Gaussian at the source position.  This template was built by averaging earlier days of observation when the source had a flux density $<$10 mJy in the same wavelength range, as measured by interferometers such as the GBI (Foster \etal 1996) or the VLA.  The flux density scale was regularly checked against the calibrator 3C433 with assumed flux densities of 12.40 and 5.60 Jy at 21 and 9 cm respectively.  Although the number of drift scans is not constant from day-to-day, the typical errors for the Nancay data set are 10 mJy (21 cm or 1414 MHz) and 20 mJy (9 cm or 3310 MHz). 

\subsection{RXTE}

The RXTE All-Sky Monitor (ASM) has been continuously observing bright X-ray sources in the range of $\sim$2 to 10 keV since 1996 February 20.  The X-ray observations of GRS 1915+105 presented herein are extracted from the ``quick-look results'' public archive provided on the World Wide Web by the ASM/RXTE team.  The light curves for both the individual 90-second exposures (typically 5-10 per day per source) and for daily average flux are available.  We have chosen to use the one-day average counts for GRS 1915+105 rather than the individual ``dwells'' as this resolution corresponds to that of our IR and radio data.  Additionally, GRS 1915+105 exhibits extreme X-ray variability over very short timescales (seconds-minutes; G96); the one day average counts smooth out these fluctuations, better representing the long timescale variations in which we are interested.  For further details about the instrument and the methods used in the ASM data reduction and error calculations, see Levine \etal (1996).

\subsection{BATSE}

The BATSE Large Area Detectors (LADs) have been used extensively and with great success to monitor hard-X-ray sources using the earth-occultation technique.  The data are sampled at 2-second intervals (although the effective time resolution here is $\sim$one 90-minute spacecraft orbit).  The useful limiting sensitivity of the earth-occultation technique is typically $\sim$100 mCrab.  For a complete description of this technique and its capabilities see Harmon \etal (1992).

We constructed a light curve spanning the interval covered by our IR campaign using the standard earth-occultation data products obtained under the auspices of a CGRO Guest Investigator Program designed to study X-ray novae outbursts.  The fluxes are derived from summations of $\sim \frac{1}{2}$ to several days, using the weighted average of typically 2-3 LADs.  In this case, count rates were converted to photon flux by assuming a power-law index of 2.8 and applying an absolute calibration.  Energy channels covering approximately 20-200 keV were included in the analysis.  

\section{Discussion}

The five three-month light curves of GRS 1915+105 we obtained are shown in Figure 1.  

\subsection{The Infrared Light Curve}

Our $K$-band light curve appears in the top panel of Figure 1.  GRS 1915+105 exhibits variability over one magnitude, from $K \sim$12.5 to $\sim$13.5.  The data show maxima at UT dates 3 July, 7 August, and 15 September.  A visual inspection of the curve between the latter two maxima indicates an apparent 40-day modulation.  An attempt to search for a regular periodicity on this timescale was unsuccessful, as a 40-day period appears inconsistent with the $\sim$30 day separation of the first two maxima.  However, we note that it is possible that our observations began {\it after} a true peak; therefore a $\sim$40-day cycle cannot be ruled out.

The shape of the curve and the long timescale for the variations are intriguingly similar to several known Be/X-ray binary (XRB) systems with eccentric orbits and long periods (generally $\gtsimeq$20 days; van den Heuvel \& Rappaport 1987).  The IR photometric characteristics of GRS 1915+105 are especially similar to the LMC Be/XRB A0538-66 (X0535-668).  A0538-66 has $P_{orb}=$16.6d and shows correlated optical and X-ray emission on this period when the system is in an active state (Skinner 1980).  Both sources show $\gtsimeq$1 magnitude variability at $K$ (M97), and both reach $L_{x}\sim10^{39}$ erg/s at outburst peak (Mirabel \etal 1996; Charles \etal 1983).  Also, similar to our GRS 1915+105 light curve, A0538-66 shows large cycle to cycle variations in its precise peak positions, mimicking a variable period.  

The inclination of the jets in GRS 1915+105 is well constrained to be $i$ = 70$\pm$2\degs (MR94).  However, this cannot be taken as a tight constraint of the {\it system} inclination; for example, the jets of GRO 1655-40 are tilted 15\degs with respect to the orbit (Orosz \& Bailyn 1997).  If GRS 1915+105 similarly has an orbital inclination tilted within $\pm$20\degs with respect to the jets, we would expect IR variability on the orbital period due to X-ray heating of the companion star and disk; however, the amplitude of these variations is partially dependent on the spectral type of the mass-donating star.  Substantial orbital flux variations are generally expected from high-$i$ LMXBs, where X-ray heating dominates the light curve.  In HMXBs, more moderate variability ($\sim$10-20\% in the optical) is the norm (van Paradijs \& McClintock 1995).  We note, however, that A0538-66 shows $>$1 mag variability at $K$ (Allen 1984).  In addition, the first maximum in our light curve is $\sim$0.3 mag brighter than the two subsequent maxima, indicating possible short-term variability such as that seen by C96.  It therefore seems likely that the IR variability in our $K$-band light curve does not result from a single cause, but from several different processes, perhaps with an underlying $\sim$40-day modulation.

In addition to orbital variability, there are a number of possible sources of IR emission in GRS 1915+105 which may account for the observed changes in magnitude.  As listed in M97, these include (1) IR emission from the accretion disk (CT96), (2) free-free emission from an X-ray driven wind (van Paradijs \etal 1994), (3) time-variable Doppler-broadened spectral line emission from ions in the relativistic jets (M97), (4) thermal dust reverberation of energetic outbursts (Mirabel \etal 1996), and (5) synchrotron emission from IR jets (Sams \etal 1996).  Of these possibilities, (3), (4), and (5) are related to ejection events and would therefore produce changes in the IR magnitude correlated with jet activity.  (Note also that the existence of the reported IR jets is in question; see Eikenberry \& Fazio 1997.)  Unfortunately, during this interval there were no high-resolution radio observations which would have provided conclusive information about ejection events which may have occurred during our observations, and discerning a clear X-ray or radio signature of such events (discussed in the following section) is problematic.  We also cannot rule out possible contamination of the IR light curve by emission from the accretion disk.  In addition, Be/XRBs exhibit an IR excess (on the order of $\sim$1 mag at $K$) associated with free-free emission in the circumstellar shell (see \eg Norton \etal 1991).  Variation would then appear during the Be star phase change (formation or dissipation of the shell).  The timescale over which such a change occurs is not known, although it is likely to be long ($\sim$months to years).  As such, it is unlikely to be the cause of the variability seen in our data.  Finally, enhanced IR emission could be produced as a result of advective accretion in the inner accretion disk (discussed in section 3.3).  

The fundamental question is, what model of GRS 1915+105 can reproduce the characterisitics of our IR light curve?  Any such model must account for the apparent independence of the IR and X-ray/radio emission from the system, as it appears that the peaks in the IR light curve are {\it not} correlated to the peaks in the radio and X-ray data (although the first two IR maxima precede increased radio emission by $\sim$10 days).  Be stars undergo erratic outbursts of equatorial mass ejection; a compact object in orbit around the Be star may therefore appear as a transient hard X-ray source, recurrent on an orbital timescale during such an outburst.  However, our light curve shows no evidence of an IR/X-ray correlation which we might expect to see as a signature of additional X-ray reprocessing of material newly accreted from the cloud during the ``close approach'' of the two components (Apparao 1985).  In and of itself, this is not sufficient to discount the Be/XRB hypothesis.  First, an X-ray periodicity correlated with increases in IR emission at periastron could easily be lost within the large, erratic X-ray activity from the disk.  Second, as in other Be/XRBs, this correlation may not always be apparent; during active states of A0538-66, the optical/X-ray correlation was {\it not} consistently observed (Howarth \etal 1984).


In fact, there is compelling IR, X-ray, and radio evidence for a circumstellar ``shell'' of gaseous material surrounding GRS 1915+105 (see the detailed discussion in Mirabel \& \rod 1996); such material is a natural consequence of Be star mass ejection.  The IR magnitude of the system increases during mass loss events, when gas is injected into the surrounding region (which may already contain remnants of previous ejections).  As the compact object sweeps through the cloud of gas near the Be star, both the shell of gas and the accretion disk of the compact object are disrupted and heated, increasing IR emission, resulting in peaks in the light curve recurrent on an orbital timescale.  Quasi-periodicity may develop as trailed material from previous passes of the compact object through the circumstellar shell may cause additional peaks in IR emission during subsequent orbits (Boyle \& Walker 1986).  A complication in this scenario is the time required for dissipation of the ejected stellar material and of the compact object's accretion disk, which may shrink as the accreted material from periastron is consumed by the black hole.  The X-ray and radio behaviour of GRS 1915+105 in our data indicate that disk activity was occurring throughout the three-month interval, \ie the accretion disk was always present.  The IR peaks could then correspond to intervals where the compact object moved through the clouds of material ejected from the Be star, which is most likely in an active outbursting state.  

Conversely, it is difficult to explain the observed IR variability with a LMXB model of the system.  In LMXBs, IR modulation is expected on the orbital period due to X-ray heating; the amplitude of this variation is generally less than 50\% in the optical, and decreases to 20-30\% in the IR (van Paradijs \& McClintock 1995).  Therefore not only is the variability larger than can be explained in a standard LMXB, the expected orbital period (usually $\ltsimeq$1 day) is substantially smaller than that seen in our light curve.  Some LMXBs do show long-term variability due to disk precession, but the amplitude of such modulation is significantly less than the 1 magnitude seen in our data (Priedhorsky \& Holt 1987).  We have not been able to reconcile the timescale of the IR variability and the amplitude of the modulation within an LMXB model for GRS 1915+105.  Therefore, we believe that the Be/XRB model, which can explain both the characteristics of the IR light curve and its apparent independence from the X-ray/radio activity, to be a more likely scenario.

\subsection{Radio and X-ray Light Curves}

\subsubsection{Description of the Data}

Radio and X-ray light curves for July-September 1996 appear in the lower four panels of Figure 1.  While the two radio curves show very similar variability, the RXTE and BATSE data are remarkably different, indicating strong spectral variations.  In the initial 15 days, erratic variability is seen in the radio and soft X-rays (RXTE).  The flux level then drops and becomes much less variable, and a period of relative quiescence ensues.  This is particularly noticeable in the RXTE data, where almost no variability is seen for approximately 20 days.  During the same 20-day period, the hard X-rays (BATSE) continue to show low-amplitude fluctuations, slowly rising throughout the soft X-ray/radio quiescence.  Then, at day 310 (15 August), a flare occurs at all four wavelengths, peaking at day 311 and then gradually falling off over $\sim$15 days in the hard X-rays and radio.  The soft X-rays continue to show large amplitude fluctuations while the flux at the other wavelengths continues to decrease.  On day 334, however, a large one-day spike in the radio flux occurs, with simultaneous smaller spikes in the X-ray flux.  The radio and hard X-ray emission then continues to decay, although the BATSE flux subsequently increases substantially (Figure 2).  The soft X-rays do not decrease simultaneously with the other wavelengths, remaining strong and variable through 1 October (day 357).  However, an examination of the complete one-year RXTE light curve of GRS 1915+105 (Figure 3) shows that the soft X-rays decayed gradually from the September peak (day 339) over the subsequent 100 days, and have returned to a low, quiescent flux level.  

\subsubsection{Comparison of the Light Curves}

Despite several radio and X-ray studies which have been performed on GRS 1915+105 and other galactic jet sources, the complex behaviour of GRS 1915+105 is not well understood, nor has a clear signature of jet ejection events been positively identified.  Foster \etal (1996; hereafter F96) found the hard X-ray (BATSE) and radio (2.25 and 8.3 GHz) emission to be generally correlated, and classified active radio emission into two states, ``plateau'' and ``flaring''.  Plateau states are characterized by a total flux density level of $\sim$100 mJy, with a duration of days to weeks.  Flaring states are always preceded by plateau states, have a very rapid rise time ($<$1 day), and are usually related to a decrease in the hard X-rays during the radio decay.  Their data showed the hard X-ray outbursts to be correlated with plateau radio states and occasionally with the occurrence of strong radio flares.  They also detected a ``plateau-flaring'' state, characterized by erratic radio emission superimposed on an underlying plateau state caused by rapid ($\leq$1 day) flares, the first time such behaviour has been seen in GRS 1915+105.  Applying these criteria to our radio (3.3 and 1.4 GHz) and BATSE data, it appears that the majority of our radio curve (days 270-338) can be described as a plateau-flaring hybrid state, similar to that seen by F96 in October-November 1995.  As expected from the F96 classification scheme, throughout this period the hard X-rays are in outburst, with BATSE flux levels consistently above 0.05 ph/cm$^{2}$/s.  We also see a strong radio flare coincident with a peak in the hard X-rays (days 310-327), similar to (but stronger than) that seen by F96 around MJD 49990, although the intense 1-day radio flare at day 334 has only a small corresponding increase in the BATSE flux.  The F96 data show one flare (10 August 1995) interpreted as an ejection event due to its rapid onset and decay time of a few days, as was detected in April 1994 (R95); this ejection event was confirmed by VLA observations (Mirabel \etal 1996).  The flare of days 310-325 in our data shares similar characteristics with both the April 1994 and August 1995 ejections.

Harmon \etal (1996; hereafter H96) have also attempted to define a hard X-ray signature for ejection events in GRS 1915+105 by examining BATSE and multi-frequency radio data.  In their examination of the behaviour of GRS 1915+105 during its 1993-94 outburst, H96 found an anti-correlation between the radio flux density and the hard X-ray flux.  This is in sharp contrast to the positive correlation seen by F96, a change which H96 attribute to the difference in radio flux levels during the two periods of activity.  During the 93-94 outburst, the radio flux was generally above $\sim$100 mJy, with extended intervals substantially in excess of $\sim$400 mJy.  In 1995, the radio flux was at a more quiescent level of $\sim$10 mJy with several episodes of $\ltsimeq$200 mJy flares.  H96 therefore assert that a positive correlation exists between radio and hard X-ray emission when the source exhibits a low GHz flux, while the weaker anti-correlation appears only at substantially higher flux densities ($\geq$200 mJy).  They suggest that this indicates more active jet production in 93-94 and differing regimes for the production of radio emission during the two epochs.  H96 surmise that the short-timescale anti-correlations correspond to ejection events (deduced from VLA interferometer data).  Our data show some evidence for an anti-correlation between the radio and hard X-rays (days 270-303); the hard X-rays are relatively low during the radio flaring of days 272-284, and subsequently increase while the radio flux drops during days 285-308.  However, in contrast to the H96 data, there is a clear correlation between the hard X-ray and radio light curves at a high flux level during the flare of days 310-327.  For the most part, our hard X-ray and radio data exhibit behaviour closer to that described by F96 than H96; as the radio flux of our data is generally lower than that seen by H96, this may support their theory that the active states are coupled to the average radio flux density at any given time, thereby producing substantially different behaviour at different epochs.

On the basis of the similarity in rise, duration, and decay times of the hard X-ray/radio flare during days 310-327 to the ejection events of April 1994 and August 1995, it seems likely that the flare in our data indicates that an ejection event occurred at this time (August 1996).  If so, it is interesting to note that the IR light curve has a maximum 8 days prior to the X-ray/radio flare.  Additionally, the first IR maximum (day 267) precedes the earlier radio flaring (possibly anti-correlated with the hard X-rays) by 7 days.  We also note that the peak of our hard X-ray light curve ($\sim$0.145 ph/cm$^{2}$/s = 480 mCrab), is higher than the peak values reported for the April 1994 ($\sim$220 mCrab) and August 1995 ($\sim$350 mCrab) ejection events.  In our data, it appears that the radio and hard X-ray peaks are coincident; however, there is a gap in our radio data between days 311 and 325.  Hjellming (1997) noted that hard X-ray peaks are followed within a few days by radio peaks in ejection events of the superluminal source GRO J1655-40, and therefore the jet ejections must begin very close to the X-ray peaks.  In the April 1994 and August 1995 ejections of GRS 1915+105, the hard X-ray flux peaks prior to the radio.  The apparent coincidence of the hard X-ray and radio peaks in our data are not inconsistent with this pattern, as it is certainly possible that the actual peak during the observed flare occurred during the gap in our radio coverage.  Therefore the pattern of the appearance of a hard X-ray peak followed within a few days by a radio peak, with subsequent correlated decays, may be a signature of jet ejection in GRS 1915+105.  It seems likely, however, that this is not the only such signature; the anti-correlations seen by H96 (also seen in Cyg X-3; Hjellming 1997) may also be a hallmark of ejection events.  As yet undefined are the mechanisms which are necessary to cause the changes in the hard X-ray/radio behaviour surrounding ejection events, and hence altering the ejection signatures of GRS 1915+105 at various times.  What does seem likely is that major ejection events do not occur if X-ray (hard or soft) active states are not accompanied by radio emission levels $\gtsimeq$100 mJy (M97; G96).

In our data, there also appears to be some correlation of the soft X-ray light curve with the radio emission.  Soft X-ray flaring occurs concurrently with the radio flaring of days 274-284; the onset of soft X-ray flaring precedes that of the radio by $\sim$9 days, and falls to quiescence $\sim$6 days before the radio flares subside.  The soft X-rays also return to outburst simultaneously with the major hard X-ray and radio flare of days 310-318, which we have labelled as an ejection event.  In the short term ($\sim$20 days after the flare), the soft X-rays continue to be high and variable while the radio and hard X-rays decline.  A strong flare coincides with the 1-day radio flare; however, the soft X-ray emission peak appears $\sim$5 days later.  As seen in Figure 3, the soft X-rays then gradually decline through the subsequent months, and finally return to quiescence.  In contrast, the hard X-rays increase to a flux level similar to that of the flare.  Examination of the BATSE and RXTE light curves from 20 February - 13 November 1996 (days 136-401) show a distinct anti-correlation of the hard and soft X-rays over the majority of that interval (Figures 2, 3); only during the suspected ejection event is similar behaviour seen.  Because of the extremely erratic variability in the radio and X-ray data, however, statistical tests of these possible correlations do not produce useful results.

F96 note that a soft X-ray outburst in March 1995 was not correlated with any detectable radio emission.  In an examination of Proportional Counter Array (PCA) data from RXTE, G96 surmise that the highly variable soft X-ray emission originates from instabilities in the accretion disk rather than from a large, hot corona.  They further assert that periods of large-amplitude X-ray variations correlated with radio emission may be related to jet formation.  This would be consistent with our observations, which show such a correlation during the onset of the possible ejection event in our data.  Origination of the soft X-ray emission in the accretion disk may also account for both its short-term variability and its continued long-term activity after the radio decay, a scenario supported by application of the advection-dominated accretion flow (ADAF) model to GRS 1915+105. 

\subsection{The ADAF Model}

The most common model for accretion flows is the thin disk model (Shakura \& Sunyaev 1973), in which heat energy released by viscous dissipation is radiated almost immediately by the accreting gas.  While this model has been successfully applied to many systems, such as accreting white dwarfs and soft X-ray transients (SXTs) in outburst, it has been unsuccessful at describing SXTs in quiescence, particularly the black hole (BH) SXTs.  However, in an advection-dominated accretion flow (ADAF) model, the radiative efficiency of the accreting gas is low, so most of the viscously dissipated energy is stored within the gas as entropy, and only a small fraction is radiated while the remainder is advected onto the compact object (Narayan \etal 1997a).  If the compact object is a black hole, then the energy disappears through the event horizon.  As a result, the total luminosity of the source will be significantly less than that expected for its mass transfer rate.  Narayan \etal (1996, 1997b) have successfully modelled the BH SXTs V404 Cyg and A0620-00 using a combination of an ADAF and thin disk model; the ADAF extends from the event horizon to an intermediate (and variable) transition radius, and a thin disk is present from this point until the outer disk edge.  The model has also been applied to Nova Mus 1991 and, most recently, the other galactic superluminal source, GRO J1655-40 (Esin \etal 1997; Hameury \etal 1997).  It therefore seems evident that this combination of ADAF/thin disk model could usefully be applied to GRS 1915+105.   

Belloni \etal (1997) have used the ADAF model to explain rapid, regular variations seen in PCA RXTE data during an active state of GRS 1915+105 in 1996.  They surmise that the emission comes in two distinct varieties: ``constant'' emission from the outer accretion disk (the thin disk), fluctuating on long timescales, while ``varying'' emission originates in the inner disk region (the ADAF), which repeatedly empties (via infall into the black hole) and refills on a timescale of seconds.  They also note that the accretion rates inferred from the X-ray spectrum are anti-correlated to the total source luminosity, as expected in an ADAF.  Finally, they suggest that the outbursts of GRS 1915+105 are caused not by an increase in the mass transfer rate, but rather by a decrease from a rate high enough to smother the X-ray emission.  In such a high state, GRS 1915+105 would only appear as a bright IR source.

We have examined our data in the ADAF/thin disk context.  If the radio and hard X-ray emission is produced close to the compact object, primarily within the ADAF region of the disk, then this emission should decay following an ejection event, as a significant fraction of the energy falls through the black hole event horizon and disappears from view.  This is borne out in our radio light curve, which clearly decreases to a level lower than seen at any other time in our data.  However, the hard X-rays decline only briefly before rising dramatically to the earlier flare level.  Any soft X-rays arising from the thin disk, \ie the ``constant'' component, could continue to be strong and variable for some time, possibly due to X-ray reprocessing.  This is consistent with the behaviour of the soft X-rays in our data both during and after the suspected ejection event.  As GRS 1915+105 is in an active state throughout our light curve, it seems unlikely that we would see the anti-correlation of the IR and X-rays expected from a ``smothered'' state.  With the multitude of other possible sources of IR emission (listed in section 3.1), at best we can only add this speculation to the list as yet another possible cause.


\section{Conclusions}

We have presented the first multi-wavelength study of GRS 1915+105 to include well-sampled IR photometry.  By examining these data, we have arrived at several conclusions.

{\bf (i)} There is evidence for a periodic IR modulation on the order of 30-40 days.  The qualitative characteristics of this variability are similar to those expected for a Be/X-ray binary in an eccentric orbit.  However, it is likely that the IR variation arises from a combination of causes, with short-term IR emission superimposed on an underlying period.  

{\bf (ii)} By comparing our X-ray/radio data to previously observed ejection events, we believe that an ejection event took place during the interval 15 August to 1 September 1996, during which time we see a large simultaneous outburst in the radio and both soft and hard X-rays.  It is interesting to note that during this event the IR emission is at a low level, indicating that IR variations resulting from jet ejection are minimal in this instance.  However, it has been suggested that the ejection event reported here actually began around day 275, when the 15 GHz radio emission was observed to increase substantially from its previously quiescent level (Pooley \& Fender 1997).  If this earlier date, which corresponds to one of the maxima in our IR light curve, is correct, then jet ejection effects may contribute to the IR flux during this interval.

{\bf (iii)} In general agreement with H96, we surmise that the pattern of a hard X-ray flare rapidly followed by a radio flare, with subsequent correlated decays, is a signature of jet ejection in GRS 1915+105; however, the X-ray activity must be accompanied by radio emission levels $\gtsimeq$100 mJy for major ejection events.  We also note that this is probably not the only such hallmark of jet ejection.

{\bf (iv)} The qualitative behaviour of GRS 1915+105 in our radio and X-ray data is in rough agreement with that expected from a two-component ADAF/thin disk accretion model.  We note, however, that as yet no detailed quantitative study of the application of this model to high-mass X-ray transients has been performed.  For example, the relative strengths of the various types of emission may be substantially different in a HMXB than in low-mass SXTs.  In addition, as GRS 1915+105 is in a high state throughout our data set, the thin disk component may dominate the X-ray and radio emission.  It is therefore difficult to draw any substantial conclusions from our data about the presence of an ADAF in GRS 1915+105.

In the near future, we will be obtaining further IR photometry of GRS 1915+105; we plan to compile a longer $K$-band light curve in an attempt to clarify the nature of the 30-40 day variability demonstrated here.  Our primary goal will be to identify the spectral type of the mass-donating star, crucial to our understanding of GRS 1915+105.  Combining an IR light curve of that length with simultaneous radio and X-ray data, as we have done here, may shed further light on the behaviour of this complex source.

\section{Acknowledgements}

The authors would like to thank the RXTE team for providing the quick-look 
results, and the BATSE team for making their data available.  The IR data 
reduction was carried out using the $\sc iraf$ and $\sc ark$ software packages 
at the Oxford $\sc starlink$ node.  RB acknowledges the support of an 
Overseas Research Scholarship.  TS was supported by a PPARC Postdoctoral 
Fellowship.

\end{document}